\documentstyle[12pt]{article}

\def\R{ {\rm R \kern -.31cm I \kern .15cm}}
\def\C{ {\rm C \kern -.15cm \vrule width.5pt \kern .12cm}}
\def\Z{ {\rm Z \kern -.27cm \angle \kern .02cm}}
\def\N{ {\rm N \kern -.26cm \vrule width.4pt \kern .10cm}}
\def\1{{\rm 1\mskip-4.5mu l} }
\def\lsim{\raise0.3ex\hbox{$<$\kern-0.75em\raise-1.1ex\hbox{$\sim$}}}
\def\gsim{\raise0.3ex\hbox{$>$\kern-0.75em\raise-1.1ex\hbox{$\sim$}}}
\def\noi{\noindent}

\def\beq{\begin{equation}}   \def\eeq{\end{equation}}
\def\bea{\begin{eqnarray}}  \def\eea{\end{eqnarray}}
\def\nn{\nonumber}
\def\noi{\noindent}
\def\beeq{\begin{eqnarray}} \def\eeeq{\end{eqnarray}}
\newcommand\mysection{\setcounter{equation}{0}\section}

\newcounter{hran}

\begin{document}
\vbox to 1 truecm {}
\centerline{\large\bf A model of quantum reduction with decoherence}

\vskip 1 truecm

\centerline{\bf Roland Omn\`es}
\centerline{Laboratoire de Physique Th\'eorique\footnote{Unit\'e Mixte de
Recherche (CNRS) UMR 8627}\footnote{e-mail: 
Roland.Omnes@th.u-psud.fr}}  \centerline{Universit\'e de Paris XI, 
B\^atiment
210, F-91405 Orsay Cedex, France}

\vskip 2 truecm

\begin{abstract}
The problem of reduction (wave packet reduction) is reexamined under
two simple conditions: Reduction is a last step
completing decoherence. It  acts in commonplace circumstances and
should  be therefore compatible with the mathematical frame of
quantum field theory and the standard model.\par

These conditions lead to an essentially unique model for
reduction. Consistency with renormalization and
time-reversal violation suggest however a primary action in the vicinity of
Planck's length. The inclusion of quantum gravity and the uniqueness of
space-time point moreover to generalized quantum theory, first
proposed by Gell-Mann and Hartle, as a convenient framework for
developing this model into a more complete theory.
\end{abstract}

\vskip 3 truecm

\noindent PCAS codes : 03.65.Ta, 03.65.Yz, 03.70.+k, 04.60.-m
 
\vskip 1 truecm

\noindent LPT Orsay 04-105\par
\noindent October 2004\par

\newpage
\pagestyle{plain}
\baselineskip 18pt

\mysection{Introduction}
\hspace*{\parindent}
Several theories or models have attempted to describe the reduction of
a wave function as a physical effect [1-6]. On the other hand, Bohm's
version of quantum mechanics in terms of ``real'' particles was
introduced to avoid reduction \cite{7r}. There is however no
universal agreement on the existence of a reduction process. Gell-Mann
and Hartle \cite{8r}, and Griffiths \cite{9r}, considered that, since quantum
mechanics is fundamentally a probabilistic theory, any mechanism
insuring the uniqueness of physical reality should necessarily stand
outside its framework. As a matter of fact, Griffiths, and the present
author \cite{10r} showed explicitly that the problem of reduction never
occurs in the framework of consistent histories, decoherence being
sufficient for all logical and practical purposes.\par

Two proposals of a more ``philosophical'' nature can also be mentioned.
As well known, Everett assumed that every quantum possibility is
actualized in some branch of a multi-valued universe \cite{11r}. An opposite
viewpoint stressed that every theory --even a classical one-- can only
deal with potentialities and never with actuality, so that the problem
of reduction might well be outside the reach of theoretical physics
\cite{12r}. Many people nevertheless consider that the problem of reduction
belongs properly to physics and should even be considered among the
most significant ones in understanding the foundations of science. \\

The discovery of decoherence \cite{13r} and its experimental confirmation
\cite{14r} have modified significantly the problem. Decoherence
explains cleanly the suppression of macroscopic quantum
interferences (except in a few well-understood cases). It is generally
followed by a classical behavior of macroscopic bodies and the various
possible results of a quantum measurement appear thereafter as a set of
well-defined possibilities obeying standard probability calculus. \par

The understanding of decoherence began with models [15-17] and was extended
to more general theories relying either on coarse graining [18-19],
predictability sieves \cite{20r} or the quantum theory of irreversible
processes \cite{21r}. Some significant problems remain however and will be
encountered in the present work. They are as follow: \par

(A). Decoherence relies on the emergence of some ``relevant'' (or
``collective'', or coarse-grained) observables, as opposed to the bulk of
all the observables entering in a more or less well-defined
``environment''.  This splitting of a macroscopic (or mesoscopic)
system into two interacting subsystems implies many drastic
consequences for the decoherent subsystem involving the relevant
observables. Its dynamic is no more unitary, it shows no quantum
interferences, and it often develops rapidly a classical behavior. The
relevant observables are thus obviously important, but their exact
meaning is still rather obscure. One can easily guess empirically what
they are, in most practical cases, but there is yet no mathematical
criterion allowing their direct construction from the basic principles
of quantum theory. \par

(B). Decoherence is often described as an
approximate diagonalization of a reduced density matrix describing
the relevant observables.  But then a problem was
pointed out by Zurek \cite{22r}: is this property general and, if then, what is
the ``pointer basis'' in which the diagonal form occurs? In the present
paper, diagonalization will not be supposed universally valid (and it is
not, as a matter of fact \cite{21r}). \par

(C). Another problem is concerned with
very small probabilities. When, for instance, decoherence amounts to
diagonalization, the non-diagonal elements of the reduced density
matrix do not vanish completely. They only decrease exponentially with
time and, in some sense, macroscopic interferences never fully 
disappear. If the initial state of the whole system is a pure state, 
it
remains so mathematically, and this survival of quantum superposition,
although through very small quantities, led some critics to consider
decoherence theory as purely phenomenological and not really
fundamental \cite{23r,24r}. Some cogent arguments have been raised however
against this point of view \cite{10r}. Anyway, it is clear that the meaning of
very small probabilities must be better understood. \par

(D). Finally,
decoherence does not of course explain or try to explain the
uniqueness of physical reality. It only implies that one can assert
consistently uniqueness according to the logic of decoherent 
histories \cite{9r,20r}.\\

The present research originated as a critique of the idea
of reduction and started from two simple conditions, which might be
expected from any theory making sense of reduction:\par

  1. If decoherence and
reduction both exist, it would be very strange if they were not
strongly related. During a measurement, decoherence appears as an
initial step destroying the quantum superposition of different results,
and reduction elects only finally a unique datum among them. Reduction
should be treated accordingly as a dynamical process, like
decoherence itself. \par

2. The uniqueness of physical reality is
obvious in the case of an
ordinary physical system involving familiar
laboratory devices. One understands quite well such an experiment, at 
least as far down as the level of particles and fields
in the standard model. Reduction, when it acts at this
commonplace level of physics, should be therefore expressible in the
mathematical formalism of well-known physics, at least in a phenomenological
way. Its action should agree with the consistency conditions of
the standard model, including relativistic invariance and
renormalization. \par

The present models of reduction disagree with these
reasonable conditions. They are always concerned for
instance with the wave functions of particles, whereas the standard
model tells us that the basic objects are quantum fields. As for
Bohmian mechanics, in spite of its elegant beginnings, it was never
able to satisfy plainly relativistic invariance and to account for
quantum fields. Conversely, the present criteria can be used in
principle for a search of possible reduction models, or for a critique
of the reduction idea.\\

Some background in decoherence and precision in the corresponding 
language will be useful for this delicate topic. Let
one assume for definiteness that the macroscopic system $S$ under
consideration can be split objectively into a relevant subsystem $R$ and
an environment $E$ (accordingly, one assumes that a solution exists for
Problem A). The relevant (or reduced) density matrix (or state
operator) is defined by

  \beq
  \label{1.1e}
  \rho_R = Tr_E(\rho_S)\ .
  \eeq
 
The coupling between $R$ and $E$ is supposed to define a set of
projection operators $\{P_j \}$ in the Hilbert space of $R$ and decoherence
leads to a block-diagonal form of $\rho_R$~:

\beq
\label{1.2e}
\rho_R = \sum_j P_j \rho_R P_j
\eeq

\noi Since the Hilbert space of $R$ is a subspace of the Hilbert space of
$S$, the operators $P_j$ are well defined as observables of $S$ (when written
as $P_{jR} \otimes I_E$, where $I_E$ is the identity operator for $E$ 
and $P_{jR}$ is a projection operator acting in the $R$ Hilbert 
space). One can then write

\beq
\label{1.3e}
\rho_S = \rho_0 + \rho_1 \ ,
\eeq
\beq
\label{1.4e}
\rho_0 = \sum_j P_j  \rho_S  P_j
\eeq

A few remarks will help understanding the meaning of these equations.
The matrix $\rho_0$  is positive with unit trace. One usually calls 
``density'' a
quantity of the same type as a density matrix (namely a trace-class
linear functional on observables, not necessarily positive or having a
unit trace), and the quantity $\rho_1$ is such a density (with zero trace).
The state of a particle or a bunch of particles not interacting
with the relevant subsystem enters in every element $P_j\rho_S P_j$ 
of $\rho_0$ as a common
tensor factor (these objects could be measured later on). The
projection operators $P_j$ and the densities are time-dependent (think for
instance of a moving part of an apparatus). Very often, they are
microlocal (semi-classical) projection operators, whose definition has
been given elsewhere \cite{12r,25r,26r}. The quantity $Tr (P_j\rho_S 
P_j)$ is a constant,
equal to the standard probability $p_j$ for observing at any time $t$ after
decoherence the property $j$ with projection operator $P_j(t)$. The density
$\rho_1$ is still poorly understood. It is clearly not zero in the case of an
overall pure state though one knows that it decreases in the trace 
norm $Tr|\rho_1 |$ when an increasing number of relevant observables 
is considered.  Its
physical meaning is related to problem A and, in the coarse-graining
approach, it represents intricate long-distance time-varying phase 
correlation in the environment. Whether or not it takes part in
reduction will not be considered in this paper.\par
 
  Another important point is the non-linearity of reduction, since the
Copenhagen theory of measurement \cite{27r}, or its derivation from first
principles \cite{12r}, shows that, when the result associated with 
the projection index $j$ is observed, the density matrix becomes

\beq
\label{1.5e}
\rho_S \to \rho '_S = {P_j  \rho_S  P_j \over Tr \left ( P_j \rho_S 
P_j\right )}\ .
\eeq

The discussion of reduction in the present paper will proceed as
follows. In Section 2, a simple model is proposed, in which the sharp
transition (\ref{1.5e}) is replaced by a continuous (dynamical) random
process, occurring after (or during) decoherence (significant analogies
between this model and a previous one by Pearle \cite{2r} should be
mentioned in that respect). Section 3 contains the core of the paper, a
theorem showing that the mathematics of reduction is essentially
unique. Quantum field theory is then introduced in Section 4, in the
simple case of a non-relativistic system of particles entering in a 
measuring device. Related projection operators are
introduced in Section 4. A discussion of infinitesimal reduction
process in Section 5 shows that it must take place primarily at
Planck's scale or nearby in order to preserve
renormalization and insure a definite time direction. Finally, some 
suggestions in Section 6 are concerned with
the theoretical framework in which reduction could enter consistently.
Generalized quantum theory, first introduced by
Gell-Mann and Hartle \cite{8r,28r}, looks particularly promising, when using
later development by Isham and coworkers \cite{29r} together with a 
recent proposal
by Kuchar on a foliation of space-time in quantized general relativity [30-31].

\mysection{A simple model}
\hspace*{\parindent}
  One is looking for a dynamical reduction process, which must therefore
proceed through infinitesimal steps. It is supposed to agree with
decoherence, so that its simplest expression is easily found as
follows. Let one write

\beq
\label{2.1e}
\rho_j = P_j \rho_S P_j\ ,
\eeq

Where as before $P_j$ is understood as $P_{jR} \otimes I_E$ and time 
is not written
explicitly. At some time $t$, an infinitesimal reduction of one state $j$
transforms the projection operator $P_j$ into $(1 + \varepsilon_j) 
P_j$. The quantum probability
$p_j = Tr\rho_j$ is then replaced by $p_j + \delta p_j$, with $\delta 
p_j = (1 + 2 \varepsilon_j)p_j$. Complete reduction consists 
ultimately in a
random process, so that $\delta p_j$ is a random quantity. It
is classically random, at least as far as one can say presently.  The
sum of all probabilities must remain however equal to 1, so that
reduction cannot affect a unique state, but in principle all of them
with the condition

\beq
\label{2.2e}
\sum_k \delta p_k  = 0
\eeq

It may be noticed that this change in the probabilities can be
represented by a change in the evolution operator $U (t)= \exp (-iHt)$. It
amounts to the replacement

\beq
\label{2.3e}
-iHdt \to - i H dt + \sum_j P_j (t) d \varepsilon_j(t) \ .
\eeq

\noi Since this change amounts formally to the use of a non-Hermitian random
Hamiltonian, it is clearly not time-reversal invariant, as one could
expect.\par

A series of infinitesimal random processes is essentially a general
Brownian variation of the coordinates $p_j$ and a few convenient
definitions will be useful. One considers only the case when the
number $n$ of coordinates is finite. The various quantities $p_k$ 
change continuously in a
random way and it is convenient to represent them as the coordinates of
a moving point $M(t)$ with coordinates $p_k(t)$ in a $n$-dimensional Euclidean
space $E$.  Since they always sum up to 1, one can introduce the
$(n-1)$-dimensional subspace $E'$, with equation $\sum p_k = 1$ in $E$. Let us
call ``vertex $k$'', or $V_k$, the point in $E$ having all its 
coordinates equal
to 0, except for the $k$-th coordinate, equal to 1. The space $E'$ contains
all the vertices $V_k$, which are the vertices of a $(n-1)$-dimensional
regular ``tetrahedron'', also often called a simplex and denoted by $S$.
The case $n = 3$ is particularly easy to visualize since the simplex is
then an equilateral triangle. Choosing an arbitrary fixed origin of
coordinates $O$ in $E'$, one has the vector relation

  \beq
  \label{2.4e}
  \overrightarrow{OM}(t) = \sum_k p_k(t)\ \overrightarrow{OV}_k\ .
  \eeq

In algebraic geometry, the quantities  $p_k (t)$ are called
barycentric coordinates of the point $M (t)$ and they can be thought
of as masses located at the vertices, $M(t)$ being their center of
gravity. They can also be considered as the distance of $M(t)$ to the $(n
- 2)$-dimensional face of the simplex opposite to the vertex $V_k$.  \par

Let $\{ \xi_j\}$ ($j = 1$ to $n - 1$) denote a set of orthogonal Cartesian
coordinates in $E'$. Their relation with the quantities $\{ p_k\}$ is simply
obtained after introducing another coordinate $\xi_0$ in $E$ along an axis
normal to $E'$, so that one has $\xi_0 = \sum p_k  = 1$ in $E'$. The 
relation between
the coordinates  $\{\xi_j , \xi_0\}$ and $\{ p_k\}$ is an $n \times 
n$  orthogonal transformation
so that, conversely,  each quantity $p_k$ is given by a first-degree
polynomial in the quantities $\{ \xi_j \}$.\par

One can draw two important consequences from this property. The fact
that the second derivatives of $p_k$ with respect to the $\xi_j$'s vanish will
play an essential role in next section. Furthermore, one can
discuss equivalently the random quantities  $d p_k (t)$ or $d\xi_j (t)$. Their
random motion will be called non-directional if the average values $<d
\xi_j (t)>$ are zero. \par

Consider then the correlation matrix $C$ whose
coefficients are defined by

\beq
\label{2.5e}
<d\xi_j d\xi_m> = C_{jm} (\xi ) dt \ .
\eeq

\noi The Brownian motion will be said homogeneous if $C$ does not depend on
$\xi$, and isotropic if it is proportional to the unit matrix. One may
notice that a condition for isotropy is invariance under the symmetries
of the simplex or, equivalently, to permutations of the quantities $d
p_k$. Because of Eq. (\ref{2.2e}), one then has

\beq
\label{2.6e}
<dp_j\ dp_k> = - {1 \over n-1} \Delta p^2\  \hbox{\rm (for $j \not= 
k)$ , and} \ <dp_kdp_k> = \Delta p^2\ , \ \hbox{for any $k$}\ .
\eeq

  The Brownian motion of the point $M(t)$, starting from the point $M 
= M(0)$, must bring it after some time on
a face of the simplex, in which one of the coordinates $p_k$ vanishes.
Thereafter, it will never go back into the simplex interior  because
the corresponding component $\rho_k$ of the density matrix has vanished. If one
now makes explicit the number $n$ in the dimension $(n - 1)$ the initial
simplex by writing it $S_n$, its boundary consists of $n$ simplexes 
$S_{n-1}$ with
dimension $n - 2$. After reaching such a face, the reduction 
mechanism becomes a Brownian
motion inside it until $M(t)$ reaches its
boundary, and the same process goes on until $M(t)$ reaches
some vertex $V_k$ in a one-dimensional simplex (which is an interval [0,
1]). Its $k$-th coordinate is then equal to 1 and the density matrix has
become $P_k\rho_S P_k/Tr(P_k\rho_SP_k)$, if one can neglect the 
density $\rho_1$ in Eq. (\ref{1.3e}). Reduction is
complete. The mechanism resulting from these simple assumptions looks
therefore able, in principle, to ``explain'' the uniqueness of physical
reality. To make sense, however, it must satisfy a very stringent
condition, which is that the probability for $M(t)$ to reach ultimately
the vertex $V_k$ should be equal to the quantum probability, $p_k (0)$,
which is the $k$-th barycentric coordinate of the initial point $M= M(0)$
from which the Brownian started.

\mysection{A uniqueness theorem}
\hspace*{\parindent}
The main problem is therefore to compute the probability $P_1$ for the
moving point $M(t)$, starting from the position $M = M(0)$ with
coordinates $(p_1(0), p_2(0), \cdots ,$ $p_n(0))$ to reach finally a
definite vertex, say $V_1$.  It may be noticed that it does not matter
whether the matrix $C$ in Eq. (\ref{2.5e}) is time-dependent or not 
as long as one does not
ask how much time the motion will take. The essential question is to find
whether there are cases yielding the quantum prediction

\beq
\label{3.1e}
P_1 = p_1(0)
\eeq

Pearle first raised this question in a different theoretical framework
and obtained insightful results \cite{1r}. The following statement 
gives a complete
answer:\\

\noi {\bf Theorem} : {\it A necessary and sufficient condition for the
validity of Eq. (\ref{3.1e}) is that the Brownian motion be
non-directional, isotropic and homogeneous.}\\

This theorem will be proved through a series of lemmas, some of them
extending previously known results [1-2, 32].\\

\noi {\bf Lemma 1} : {\it Eq. (\ref{3.1e}) is valid for a 
non-directional, isotropic and
homogeneous Brownian motion.}\\

\noi {\bf Proof} : Let one first show that the function $P_1(p_1, 
\cdots , p_N)$ is harmonic in the $\xi$
variables, the initial value of time $(0)$ characterizing these
quantities being omitted, just like $M(0)$ was denoted by $M$. One 
knows already that this
function is harmonic if Eq. (\ref{3.1e}) holds, since $p_1(0)$ has 
vanishing second derivatives in the $\xi$ variables. Conversely, a 
function is
harmonic if and only if its value at any point $M$ is the average of 
its values over
an arbitrary $(n - 2)$-dimensional sphere centered at $M$. Let one 
then consider a sphere $\Sigma$ (with element of area $d\Omega$), 
centered
at $M$ and inside the open simplex $S_n$. Denoting by $Q$ an arbitrary point on
$\Sigma$, the harmonic character of $P_1(p_1, \cdots , p_N)$  will be 
established if one has

\beq
\label{3.2e}
P_1(M) = \Omega^{-1} \int_{\Sigma} P_1(Q) d \Omega \ .
\eeq

This is easily shown as follows: The moving point $M(t)$ starting from $M$
must cross the sphere $\Sigma$ before reaching finally  vertex 1. Let 
$\Pi (M, Q)d\Omega$
denote the probability for reaching $\Sigma$  for the first time in 
an infinitesimal region $d\Omega$ containing $Q$. Using composite 
probabilities, one has:

\beq
\label{3.3e}
P_1(M) = \int_{\Sigma} \Pi (M, Q) P_1(Q) d\Omega \ .
\eeq
 
\noi But for an isotropic Brownian motion, one has $\Pi (M, Q) = 
\Omega^{-1}$, so that Eq. (\ref{3.2e})
is true.\par

The next step consists in showing that $P_1(p_1, \cdots , p_N)  = 
p_1$. This is obvious when $N =
2$, when the simplex reduces to an interval (for instance the interval
$V_1V_2$) and there is only one coordinate $\xi_1$. $P_1$ depends 
then only on the
variable $p_1$, which is a first-degree polynomial in  $\xi_1$. Being harmonic,
$P_1$ is a function $A p_1+B$, which is furthermore equal to 1 when the
starting point is the vertex $V_1$  (where $p_1 =1$) and equal to zero at $V_2$
(where $p_1 =0$). Therefore $P_1 = p_1$.\par

When $n = 3$, the simplex $S_3$ is an equilateral triangle. The case $n = 2$
has already shown that the boundary value of the harmonic function 
$P_1(p_1,p_2,p_3)$  is
equal to $p_1$ on both sides $V_1V_2$  and $V_1V_3$ of the triangle 
and it obviously
vanishes on the side $V_2V_3$ (where $p_1 =0$). The expected result 
follows then
from the uniqueness of the solution of the Dirichlet problem. The
extension to any value of $n$ by means of recurrence is obvious. \\

\noi {\bf Lemma 2} : {\it Eq. (\ref{3.1e}) does not hold for an 
anisotropic homogeneous Brownian motion.} \\

\noi {\bf Proof} : The matrix $C$ is then a symmetric positive 
matrix, with some
different eigenvalues. One can choose the axes of coordinates along its
eigenvectors and perform a change of scale $\xi_j \to \eta_j = 
\lambda_j \xi_j$ bringing $C$ proportional to
the unit matrix.  The simplex is no more regular in the $\eta$ coordinates,
but the previous reasoning implying the harmonic character of $P_1$ (in $\eta$)
remains valid.  The previous result also holds true for $n = 2$, since a
Brownian motion is always isotropic in one dimension. But if one tries
to descend from a simplex $S_n \cdot  (n \geq 2)$ to its boundary, 
the boundary value
of $P_1$, when expressed as a function of the $\eta$ variables of a 
boundary simplex
will be harmonic only if the variables of
type $\eta$ agree on both simplexes $S_n$ and $S_{n-1}$. But this 
property holds only when
the boundary simplex $S_{n-1}$ is spanned by $n-2$ eigenvectors of 
$C$ and it cannot
hold for all the boundary simplexes. In other words, the boundary
value of $P_1$ cannot be harmonic everywhere on the boundary, forbidding
the descent process. Another way of getting the same answer consists in
looking at the transfer from $n = 2$ to $n = 3$. The values of $P_1$  resulting
from the one-dimensional Brownian motions on the three sides of the
equilateral triangle have not the same algebraic expression in the $\eta$
variables.\\

\noi {\bf Lemma 3} : {\it Eq. (\ref{3.1e}) does not hold for an 
inhomogeneous Brownian motion.}\\

\noi {\bf Proof} : In view of Lemma 2, one may assume the motion isotropic.
It will be enough to establish Lemma 3 in the case $n = 2$, since the
property (\ref{3.1e}) cannot hold for an arbitrary value of $n$ if it does not
hold for the final one-dimensional step for which $n = 2$. Let one then choose
the coordinate $\xi = p_1$, defined in the interval $J = [0, 1]$. One 
can use the
diffusion approximation for an inhomogeneous Brownian motion, where the
probability distribution of the moving point $M(t)$ is a function 
$\rho (\xi , t)$
satisfying the following equation

\beq
\label{3.4e}
\partial \rho /\partial t = {\partial \over \partial \xi} \left \{ 
D(\xi ) {\partial \rho \over \partial \xi} \right \} \ ,
\eeq

\noi where the diffusion coefficient $D(\xi)$ is positive.\par

The boundary conditions for absorbing boundaries are given by

  \beq
  \label{3.5e}
  \rho (0, t) = \rho (1, t) = 0\ .
  \eeq

The initial value is taken as

\beq
\label{3.6e}
\rho (\xi , 0) = \delta (\xi - \alpha ) \ ,
\eeq

\noi where $\alpha$ is the initial position $\xi (0)$ of the moving point.\par

Denoting by $L$ the operator $\partial / \partial \xi (D\partial / 
\partial \xi)$), one introduces orthonormal
eigenfunctions and eigenvalues  satisfying the differential equations

\beq
\label{3.7e}
L\psi_n = - \lambda_n \psi_n\ ,
\eeq

\noi with Dirichlet boundary conditions $\psi_n(0) = \psi_n(1) = 0$. 
The quantities $\lambda_n$  are
strictly positive (zero is not an eigenvalue).\par

The delta function in Eq. (\ref{3.6e}) is written as

\beq
\label{3.8e}
\delta (\xi - \alpha ) = \sum_n \psi_n (\xi ) \psi_n (\alpha ) \ ,
\eeq

\noi so that

\beq
\label{3.9e}
\rho (\xi , t) = \sum_n \psi_n (\alpha ) \psi_n (\xi ) \exp (-\lambda_n t) \ .
\eeq
 
The probability that the point $M(t)$ ends up at the boundary $\xi  = 1$ is
given by the time integral of the flux

  \beq
  \label{3.10e}
  P_1 = - D(1) \int_0^{\infty} {\partial \rho (1, t) \over \partial 
\xi } dt = - D (1) \psi_n (\alpha ) \psi '_n (1)/\lambda_n\ .
  \eeq

The right-hand side of this expression has a simple interpretation after
introducing the Green function (with complex $z$).
 
  \beq
  \label{3.11e}
  G(z, \xi , \eta ) = \langle\xi | (L-zI)^{-1}|\eta \rangle \ ,
  \eeq

\noi i.e. the kernel of the operator $z\ I - L$ with Dirichlet 
boundary conditions. One has

  \beq
  \label{3.12e}
  \left . P_1 = D(1) {\partial \over \partial \xi} \ G(0, \alpha , \xi 
) \right |_{\xi=1}\ .
  \eeq
 
The limit $\xi  = 1$ in the derivative is meaningful, because $G(0, 
\alpha  , \xi)$
is an infinitely differentiable function of $\alpha$  and  $\xi$ 
(except for $\alpha  =
\xi$ ) in the Cartesian product $J \times J$. Considering  $P_1$  as 
a function of $\alpha$ and
letting the operator $L$ act on it, one gets

  \beq
  \label{3.13e}
  D(\alpha ) {d^2P_1 \over d\alpha^2} + D'(\alpha ) {dP_1 \over d 
\alpha} = D(1) \delta '(\alpha - 1) \ .
  \eeq
 
(Note : A rigorous use of the delta function derivative at the point $\xi = 1$
can be justified through a limiting process $\xi \to 1$, but this is trivial
because the only property we need is $\delta ' (\alpha - 1)= 0$, 
when $\alpha  < 1$. One thus gets

  \beq
  \label{3.14e}
  {d \over d \alpha} \left ( D(\alpha ) {dP_1 \over d \alpha} \right ) 
= 0 \ , \hbox{or}\ {dP_1 \over d \alpha} = {C \over D(\alpha )}\ ,
  \eeq

\noi where $C$ is a constant.  Eq. (\ref{3.1e}), which would suppose 
$P_1 (\alpha ) = \alpha$, holds
therefore only when $D(\alpha)$ is a constant, i.e. only in the case of
homogeneous Brownian  motion.\\

\noi {\bf Lemma 4} : {\it Eq. (\ref{3.1e}) does not hold for a 
directional Brownian motion.}\\

\noi {\bf Proof} : Consider the simplest case $n =2$, with $<d\xi /dt> = \nu$.
In the diffusion approximation, the probability distribution satisfies
the Fokker-Planck equation

\beq
\label{3.15e}
\partial \rho / \partial t = {\partial \over \partial \xi} \left ( 
D{\partial \rho \over \partial \xi} - \nu \rho \right ) \ .
\eeq

\noi One can integrate it in time from zero to infinity, using the initial
condition (\ref{3.6e}), to get

\beq
\label{3.16e}
- \delta (\xi - \alpha ) = Dd^2y/d\xi^2 - \nu dy/d\xi\ , \hbox{with}\ 
y(\xi ) = \int_0^{\infty} \rho (\xi ; t) dt \ .
\eeq

This elementary differential equation determines explicitly the
function $y$ in view of the boundary conditions (\ref{3.5e})~:

$$y(\xi) = {D \over \nu}\ {e^{\nu\alpha/D} - e^{\nu/D} \over 
e^{\nu/D}-1} \left ( e^{\xi \nu/D} - 1 \right ) \vartheta (\alpha - 
\xi ) - {D \over \nu} \cdot {1 - e^{-\nu\alpha/D} \over e^{\nu/D}-1} 
\left ( e^{\xi\nu/D} - e^{\nu/D}\right ) \vartheta (\xi - \alpha )\ 
,$$

\noi where $\vartheta$ denotes the unit step-function. Noticing that
$P_1 = - Dy'(1)$, one thus gets the exact expression

  \beq
  \label{3.17e}
  P_1 = \left ( 1 - \exp (- \nu \alpha /D)\right )/\left ( \exp 
(\nu/D)-1\right ) \ ,
  \eeq

\noi which satisfies condition (\ref{3.1e}) when and only when $\nu = 0$.\\

Finally, one can estimate roughly the time during which a complete
reduction occurs when starting from $n$ competing states. A
unique scale of time $\tau$ will be assumed to enter in  the average 
displacement of
the moving point $M(t)$, whatever the value of $n$:

\beq
\label{3.18e}
\Delta \xi^2 = t/\tau\ .
\eeq

At every step, or more properly during every step of the descending
scale in $n$, the point $M(t)$ has moved
over a distance $\Delta \xi$ of the order of unity  (the size of
the simplex) when it reaches a boundary. Accordingly, the duration of a
complete reduction is of the order of $n \tau$ . Estimating the value of
$\tau$ remains however an open problem.

\mysection{Non-relativistic systems and quantum field\break\noindent theory}
\hspace*{\parindent}
It was assumed in the introduction that the action of reduction 
should be expressible in the framework of quantum field
theory, at least in a phenomenological way. Actually, the
uniqueness of physical reality is usually observed in commonplace
occasions,  under
non-relativistic conditions. One may therefore begin by considering a 
non-relativistic macroscopic
system. Elementary quantum mechanics considers it
as made of particles, though one knows that quantum fields are more
fundamental. As a first task, one must therefore cast the description of
a system of non-relativistic particles into the framework of quantum
field theory.\par

One considers for convenience a simple model of a macroscopic system
where all the particles belong to the same species and are described by
a scalar relativistic field $\phi (r, t)$. In the reference frame 
where the system
is at rest, one can introduce a field $\Pi (r, t)$, canonically 
conjugate to $\phi (r, t)$.
The evolution amplitudes $<\phi_2t_2|\phi_1,t_1>$ are given by  Feynman sums

  \beq
  \label{4.1e}
  \int [d\phi ] [d\Pi] \exp \left \{ i \int dt \int (dr) \left [ L + 
\phi (x) \rho (x) + \Pi (x) \sigma (x) \right ] \right . \ .
  \eeq

\noi The notation for Feynman sums and ordinary integrals, which is 
borrowed from
Brown's book \cite{33r}, distinguishes the space integration in the 
Lagrangian, denoted by $(dr)$ from a Feynman summation, denoted with 
square
parentheses $[d\phi]$. The four-dimensional variable $x$ stands for $(r,
t)$. The Lagrangian density $L$ is given by $\Pi \partial_0\phi - 
H(\phi , \Pi )$, and includes the Hamiltonian
$H$. Sources $\rho$ and $\sigma$ have been introduced for later 
convenience. The
two fields $\phi$ and $\Pi$  satisfy the canonical commutation relations

\beq
\label{4.2e}
\left [ \phi (r, t) , \Pi (r', t)\right ] = i \delta^{(3)} (r-r')\ ,
\eeq

\noi and the same notation is used for the quantum fields and the
classical fields on which Feynman sums are performed. Free fields are
characterized by the Hamiltonian
 
  \beq
  \label{4.3e}
  H = {1 \over 2} \Pi^2 + {1 \over 2} (\nabla \phi )^2 + {1 \over 2} 
m^2\phi^2 \ .
  \eeq
 
A non-relativistic version of the field formalism is obtained when mass
terms $e^{\pm imt}$  are separated from the phase of the fields, 
thereby introducing a pair of
non-relativistic fields as

  \bea
  \label{4.4e}
  &&\psi (r, t) = e^{imt} \left [ \sqrt{(m/2)} \ \phi (r, t) + i / 
\sqrt{2m}\ \Pi (r, t)\right ] \ , \nn \\
  &&\psi^{\dagger} (r, t) = e^{-imt} \left [ \sqrt{(m/2)} \ \phi (r, 
t) - i / \sqrt{2m}\ \Pi (r, t)\right ] \ .
  \eea

\noi The commutation relations of these fields are those of local creation and
annihilation operators, when the quantities $e^{\pm2imt}$ are 
considered as rapidly varying and averaged
out:

\beq
\label{4.5e}
\left [ \psi (r, t) , \psi^{\dagger} (r', t)\right ] = \delta^{(3)} (r-r')\ .
\eeq

\noi Neglecting similar rapidly changing terms in the action, the Feynman
sum (\ref{4.1e}) for a free field becomes in the non-relativistic limit:
 
\beq
\label{4.6e}
\int [ d\psi ] [d\psi^*] \exp \left \{ i \int dt \int (dr) \left 
[\psi^* i {\partial \psi \over \partial t} - \psi^* \left ( - 
{\nabla^2 \over 2m}\right ) \psi + \psi^* f + \psi g)\right ]\right 
.\ .
\eeq

\noi The sources $f$ and $g$ are linear
functions of $(\rho, \sigma)$. The effect of an external potential 
$V_1(r)$  and
of interaction potentials $V_2(r,r')$ results from adding the
following quantity to the Lagrangian:

\beq
\label{4.7e}
\int (dr) \psi^* (r, t) V_1(r) \psi (r, t) + {1 \over 2} \int (dr) 
(dr') \psi^* (r,t) \psi^*(r',t) V_2(r,r') \psi (r,t) \psi (r',t)\ .
\eeq

\noi Finally, the relation with particle wave functions is obtained 
from introducing a
vacuum state $|0>$, so that a pure state of the ssytem with wave 
function $u$ is given in the field formalism by

\beq
\label{4.8e}
|u> = \int dr_1\cdots dr_n u(r_1, \cdots r_n, t) 
\psi^{\dagger}(r_1,t) \cdots \psi^{\dagger} (r_1, t)|0>\ .
\eeq

\mysection{Infinitesimal reduction}
\hspace*{\parindent} One now comes to the central problem, which is 
the compatibility of a
reduction mechanism with known physics and, particularly, relativistic
quantum field theory as it is used in the standard model. The
difficulty underlying this problem is emphasized by the fact that no
previous model of reduction was able to solve it. The root of the
difficulty, in my opinion, does not lie really in the possibility of
dealing with a macroscopic system in an arbitary reference frame. There
is no reason to expect this problem of relativistic invariance to be
more difficult than in standard quantum field theory. The real
difficulty may well come from the projection operators entering in
reduction, whether continuous or sudden.\par

Since the introduction of reduction by Bohr, and later in the models
attempting to formalize it, it was always assumed that a specific part
$u_{\lambda}$ of the wave function $u$ (representing a definite
measurement result $\lambda$) was preserved, while the rest of the wave
function vanished. It will be instructive to look at the projection
operator $P_{\lambda}$ realizing this effect, either completely or
infinitesimally. \par

In the non-relativistic quantum field formalism of the previous
section, this projection operator can be expressed in terms of fields
by

\bea
\label{5.1e}
&&\int (dr_1) \cdots (dr_n) \cdot dr'_1 \cdots dr'_n 
\psi^{\dagger}(r_1) \cdots \psi^{\dagger}(r_n) \cdot \psi (r'_1) 
\cdots \psi (r'_n)\nn \\
&&\times u_{\lambda}(r'_1, \cdots , r'_n) u_{\lambda}^*(r_1, \cdots , r_n) \ .
\eea

\noi This expression cannot be extended however to a less trivial field
theory. In quantum electrodynamics for instance, the Hilbert space
involves photons, particles and antiparticles. There is no wave
fucntion describing the particles as such, but a state involving the
values of the various fields at every point of space. The operator
$P_{\lambda}$ could be written in the form (\ref{5.1e}) because the
$(\psi, \psi^{\dagger})$ fields destroy and create particles, whereas
no operator can destroy or create the fields themselves. It seems
therefore that one cannot express in quantum field theory a reduction
process that would act on every detail of a wave function. I am
inclined to believe that this simple remark is the main reason why
the GRW mechanism or Bohm's model could not be extended to field
theory.\par

The situation is quite different when decoherence has acted first. It
was shown already that the projection operators entering in reduction
are much simpler and this point can be made clear on a simple example.
Consider a case where the relevant subsystem $R$ consists of a unique
massive particle with position  $r_0$ and field $\psi_0(r_0, t)$. The
environment $E$ consists of a large number of identical particles, as
in Section 4 (this model being quite similar to Joos-Zeh's model
\cite{17r}, if the massive ``particle'' is supposed to have a spatial
extension). Let one assume that decoherence has split the state into
mutually exclusive events $\lambda$, each one of them corresponding to
the location of the massive particles in a different region of space
$D_{\lambda}$. The corresponding projection operator (which, $E_E \to 
I_E$ was denoted
earlier by $P_{\lambda R} \otimes E_E)$ is simply given by

\beq
  \label{5.2e}
  \int_{D_{\lambda}} dr\psi_0^{\dagger}(r) \psi_0 (r) \ .
  \eeq

Although I will not discuss the extension of this formula to a less
trivial field theory, one does not expect any specific difficulty,
since this operator belongs to the field algebra. In any case, the
present considerations are only preliminary and they are only meant as
a first encounter with past difficulties and their expected removal.\par

Finally, one must consider more significant consequences of the 
consistency of reduction with the standard model. Reduction violates
time-reversal invariance and there are strong indications from the
standard model that interactions violating time reversal
become sizable at the scale of Planck's
length. Another remark points also to a fundamental action of reduction
at this level: When a projection operator is integrated over
time with a random factor $d\varepsilon (t)$, it behaves formally as 
a source term analogous
to the ones in Eq. (\ref{4.1e}). It involves however generally a 
somewhat complicated function of the fields.
  Such
source terms will obviously break renormalization, except if 
$d\varepsilon (t)$ varies
  rapidly.  It is generally
agreed that this situation can happen only when the time scale of 
variation of the
source is of the order of Planck's time.
 
\mysection{Perspectives}
\hspace*{\parindent}
Having to deal together with a source of reduction at the scale of 
Planck's length and the
macroscopic uniqueness of space-time raise immediately the
questions of quantum gravity and cosmology. I shall restrict however 
the remarks
on this topic to the perspective of generalized quantum theory, which
was introduced by Gell-Mann and Hartle for this special purpose, as an
extension of consistent decohering histories \cite{8r,28r}. Isham and coworkers
devised for it a convenient mathematical formalism.
Rather than using a unique Hilbert space, they introduced a tensor
product of as many copies of the Hilbert space as there are
instants of time in a family of consistent histories \cite{29r}. The
procedure relies however on a foliation of space-time by
space-like surfaces and the existence of such a foliation looked
questionable in the Wheeler-De Witt theory of quantized
gravity, in which there is no time variable. A recent proposal by
Kuchar might open new possibilities for this approach and will be
described here briefly \cite{31r,32r}. \par

In the canonical formalism of quantum mechanics, a unique Hilbert space
provides a representation of the commutation relations between
canonically conjugate operators (for particles or fields) $[x, p] = 1$.
Isham's formalism involves a set of time-indexed operators satisfying
history-dependent commutation relations, which can be written
essentially as $[x(t), p(t')] =i\delta (t-t')$. Kuchar's proposal consists in
extending the space-time phase space by introducing a time operator 
$T$ together
with its conjugate momentum. In the case of vacuum gravity, the corresponding
``phase space'' involves then a set of operators

  \beq
  \label{6.1e}
  \left ( G_{AB}(y) , \Pi^{AB}(y);T(y),\Pi_T(y)\right )
  \eeq

\noi and a key point is that these space-time fields can be freely varied
in the space-time classical action. Projection operators, as they enter
in consistent histories, can also be defined.

Kuchar considered the case of pure gravity, but a very interesting
problem will occur when the time foliation will be coupled to
decoherence. The consistency of decohering histories will strongly
constrain the time foliation and reduction might give it an objective
local meaning. This is of course only a speculative perspective, but it
looks encouraging for further investigations.

Also encouraging is the convergence of older approaches to the
problem of reduction with the present one. Although it was described
here in a constructive way, the present approach was initially meant as a
critical analysis of the previous ones, according to the two basic
conditions that were stated in the introduction.  It looks remarkable 
that a critical approach brought thus back rather naturally new 
variants of
spontaneous reduction \cite{3r,4r}, stochastic continuity 
\cite{1r,2r} and of the
relevance of quantum gravity in the uniqueness of space-time \cite{5r,6r}.

\section*{Acknowledgements}

I thank Bernard d'Espagnat for useful remarks and Karel Kuchar for an 
illuminating discussion on related topics. I also thank the 
organizers of the Second DICE meeting in Piombino, and some speakers 
in whose lectures I found an inspiration for some of the ideas 
occurring in this work.

\end{document}